\documentclass[aps,prd,reprint,amsmath,amssymb, notitlepage, twocolumn,
nofootinbib,
superscriptaddress,
longbibliography
]{revtex4-1}

\usepackage{amsmath}
\usepackage{tabularx,graphicx}
\usepackage{epstopdf}
\usepackage{graphicx}
\usepackage{latexsym}
\usepackage{amssymb}
\usepackage{amsmath}
\usepackage{color, colortbl}
\usepackage{psfrag}
\usepackage{bbm}
\usepackage{bm}
\usepackage{titlesec}
\usepackage{dsfont}
\usepackage{feynmp}
\usepackage{slashed}
\usepackage{multirow}
\usepackage{url}

\newcommand{\beq}{\begin{eqnarray}}
	\newcommand{\eeq}{\end{eqnarray}}

\newcommand{\bsp}{\begin{split}}
	\newcommand{\esp}{\end{split}}

\newcommand{\ie}{{i.e., }}
\newcommand{\eg}{{e.g., }}

\usepackage{color}
\definecolor{darkblue}{rgb}{0.,0.,0.4}
\definecolor{darkred}{rgb}{0.5,0.,0.}
\definecolor{BlueViolet}{RGB}{138,43,226}
\definecolor{SkyBlue}{RGB}{30,144,255}
\definecolor{DarkGreen}{RGB}{0,100,0}
\usepackage[pdftex,colorlinks=true,linkcolor=darkblue,citecolor=blue,urlcolor=darkred]{hyperref}
\usepackage[normalem]{ulem}

\usepackage{mathrsfs}
\usepackage{comment}

\begin{document}

\title{Renormalization-group Perspective on Gravitational Critical Collapse}

\author{Huan Yang}
\email{hyang@tsinghua.edu.cn}
\affiliation{Department of Astronomy, Tsinghua University, Beijing 100084, China}
\affiliation{Perimeter Institute for Theoretical Physics, Ontario, N2L 2Y5, Canada}
\affiliation{University of Guelph, Guelph, Ontario N1G 2W1, Canada}

\author{Liujun Zou}
\email{zou@perimeterinstitute.ca}
\affiliation{Perimeter Institute for Theoretical Physics, Ontario, N2L 2Y5, Canada}

\begin{abstract}

In this work, 
we propose extremal black holes (BH) as critical points of a new class of gravitational collapses. The conjecture is made by observing the Continuous Self Similarity (CSS) and Discrete Self Similarity (DSS) behaviours of perturbations of an extremal black hole spacetime and compare them to similar properties of Choptuik-type critical solutions. By performing analytical perturbation studies on extremal black holes, we explicitly show that the DSS solution found here can be interpreted as renormalization group (RG) limit cycles, and the transition between CSS and DSS regimes occurs as the stable and unstable fixed points collide and move to the complex plane.  We argue that the DSS solutions found in spherically symmetric gravitational collapses can be similarly interpreted. We identify various phenomena in non-gravitational systems with RG limit cycles, including DSS correlation function, DSS scaling laws in correlation length and order parameters, which are observed in gravitational critical collapses. We also discuss a version of gravitational Efimov effect.

\end{abstract}

\maketitle

\section{ Introduction}

Gravitational critical collapse (GCC), initially discovered by Choptuik in 1993 \cite{Choptuik:1992jv}, represents a class of the most extensively studied gravitational critical phenomena in the strong-gravity regime. Near the collapse critical point, the spacetime and the field(s) may exhibit self similar behavior in certain parameter regime, which is often referred to as Continuous Self Similarity (CSS). In other cases fields and the spacetime may instead display oscillatory, periodic behaviour by varying the scale of the system, thus breaking the continuous self similarity of the configuration. Such phenomena are often referred to as Discrete Self Similarity (DSS). Away from criticality, the system's order parameter $M$, e.g., the mass of black hole formed after the gravitational collapse, is often related to a control parameter $p$, which characterizes the profile of initial data, by $M-M_\star \propto (p-p_\star)^\gamma$ ($\gamma$ is a critical exponent). Such behaviour is very similar to that in condensed matter systems near continuous phase transitions.

Indeed the classification of GCC follows closely with the analogy of phase transitions, as nicely reviewed in \cite{Gundlach:2007gc}. Various kinds of collapse simulations, with different underlying theories, spacetime dimensions, initial data and evolution schemes, have been performed in the past decades (Table I in \cite{Gundlach:2007gc}), which greatly enrich the phenomenology of GCC and point to deeper connection with condensed matter systems near criticality. It is also understood how to compute the critical exponent and the DSS period in certain systems with eigenvalue analysis of wave equations (e.g., \cite{Gundlach:1996eg,Martin-Garcia:1998zqj,Husa:2000kr,Koike:1995jm}). However, despite this great progress, many fundamental questions remain unsolved. For example, how ``universal" (i.e., whether they change among different theories and/or model parameters) are the critical exponent and DSS period, and how are they connected to/determined by the critical point? Why is there a DSS behaviour in many GCC experiments? Is there a more explicit relation/mapping between GCC and condensed matter systems near criticality? If so, is there a unified framework to describe critical behaviours in both gravitational and non-gravitational systems?

While the theoretical impact of answering these questions is  profound, it requires a community effort to eventually obtain answers, which may benefit from the analysis on analytically solvable systems in addition to numerical simulations. In this sense, studying new critical phenomenon should also help make progress in this direction. In this work, we propose that extremal black holes are the critical points of a new class of GCC, and discuss possible setups to realize the critical process.  Although no such numerical implementation has been performed so far, there are analytical arguments supporting this claim.
In addition,  this viewpoint, i.e., extremal black holes as critical points of GCCs, provides an analytically solvable system to understand various properties of GCC, and its connection to critical behaviour in non-gravitational systems.
In particular, we show that the extremal black hole collapse experiments can be reformulated via RG. It will be interesting to check whether such viewpoint  also extends to generic GCC experiments.
In fact, it is known that the correlation function in quantum systems with RG limit cycles (fixed points) displays DSS (CSS) behaviour, which shows interesting analogy with the Green's function in GCCs. 
In addition, quantum systems with RG limit cycles near the critical point naturally requires that the order parameter (and the correlation length) scale as
\begin{align}
M-M_\star \propto (p-p_\star)^\gamma e^{f[\log (p-p_\star)]}
\end{align}
where $f(.)$ is a periodic function with certain period $\Delta$ and the scale of the system remains fixed as we tune the control parameter $p$. This DSS behaviour of order parameters is also generically seen in GCC experiments near DSS critical points. 
In the gravity side, by comparing the Choptuik-type collapse experiments and the proposed experiments with extremal black holes as the critical points, we conjecture that in certain collapse experiments, there are extremal-black-hole-like singularities with the continuous conformal symmetry explicitly broken, but nevertheless satisfies discrete conformal symmetry, i.e., the spacetime is DSS.

The  RG  interpretation of GCC calls for more quantitative mappings between gravitational and quantum (or statistical physics) systems.
To establish such mappings, observable like the critical exponents, the correlation/Green function may be useful, and we show that the DSS period $\Delta$ is not suitable for this purpose as it is non-universal, as it changes for different model parameters (see discussion in Sec.~\ref{sec:extp}).
Much is unknown along this direction and we shall discuss a few open problems. For simplicity we adopt the natural unit system where $c=G=1$.

\vspace{0.2cm}
\section{Extremal black holes from GCC.}

In this section, we discuss extremal black holes and their perturbations in two aspects. In Sec.~\ref{sec:extp} we point out that the perturbations of extremal black holes share similar CSS and DSS behavior as Choptuik-like critical solutions, which motivates the conjecture that extremal black holes are critical solutions of a new type of GCC. \footnote{A rigorous proof appeared on arXiv \cite{Kehle:2024vyt} during the preparation of an updated version of this manuscript. Our analysis nevertheless provides a different perspective on why forming extremal black holes from GCC is expected.} On the other hand, in Sec.~\ref{sec:rgp} we perform renormalization group analysis to the wave equation to explain why both DSS and CSS signatures may appear for the critical solutions and/or their perturbations. In particular, we point out the DSS behavior is related to the renormalization group limit cycles.

\begin{figure}[h]
\centering
\includegraphics[width=0.36\textwidth]{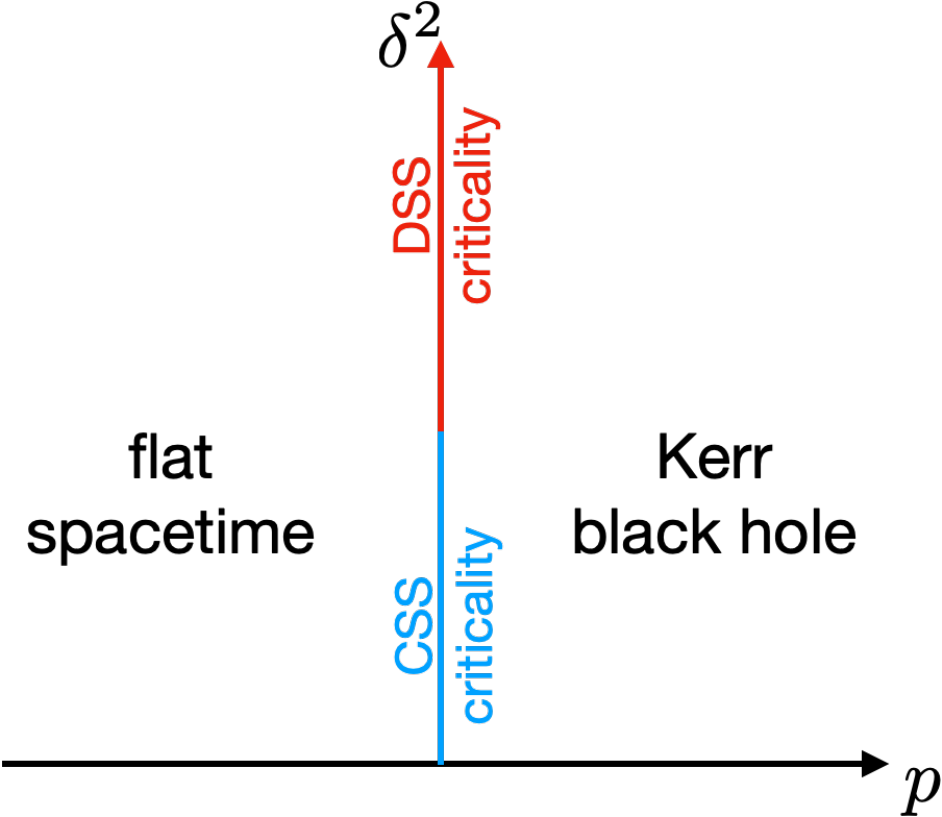}
\caption{The schematic phase diagram, with the horizontal axis the initial data that tunes the transition between a flat spacetime and a Kerr black hole, separated by a vertical axis denoting an extremal Kerr spacetime. Depending on the sign of $\delta^2$ the perturbations of the critical solution may display either continuous or discrete self similar behavior.}
\label{fig: phase diagram}
\end{figure}

With non-spherical perturbations applied to the Choptuik-like collapse experiments, it has been suggested that the spherically symmetric critical solution is stable \cite{Martin-Garcia:1998zqj}. Intuitively to avoid this type of critical solution, we can start with spherically symmetric initial data (scalar field or fluid) deep in the black hole formation regime, and then gradually increase the angular momentum of the initial data. The resulting end state should be a black hole with increasing spin.  Since a non-extremal
black hole is able to spin up by accretion, we expect the maximum spin achievable in this process is $a/M=1$ ($a$ is the reduced spin), i.e., an extremal Kerr, unless there is additional critical process associated with matter present in the collapse process. Notice that according to the Weak Cosmic Censorship, a rotating black hole cannot have super-critical spin from gravitational collapses, with matter satisfying the null energy condition \cite{Sorce:2017dst}. This means that the only viable alternative outcome of this type of gravitational collapse, as the angular momentum of initial data continues to increase, is a flat spacetime with waves/matter dispersed away. The extremal Kerr may serve as the critical point for this type of GCC (see Fig. \ref{fig: phase diagram}). Similarly, extremal Reissner-Nordstrom may serve as the critical point for a spherically symmetric collapse with charged fields. The collapse of spherical charged dust analyzed in \cite{Boulware:1973tlq} indeed shows that the dust bounces and disperses to infinity if $Q>M>M_0$, with $Q$ being the total charge of the dust shell, $M$ being the total mass and $M_0$ being the rest mass, and collapses to be a Reissner-Nordstrom black hole if $M>Q>M_0$.

To conveniently probe the critical behaviour, one may perform GCC experiments with either rotating or charged waves or matter. The analysis in Sec.~\ref{sec:extp} not only motivates the existence of such critical solution, but also provides theoretical predictions on the critical exponent, the possible DSS period, etc., which can be further tested by the numerical experiments.


\vspace{0.2cm}
\subsection{ Extremal black holes and Perturbations.}\label{sec:extp}
In this section, we discuss the critical signatures of perturbative fields on a background extremal black hole. This is motivated by a similar discussion of Choptuik-type of gravitational critical collapses, where DSS and/or CSS critical solutions and perturbations are found and used to characterize the critical collapse process \cite{Gundlach:1996eg}. For simplicity we shall analyze the perturbation of scalar waves. Since the Teukolsky equation for fields with different spin weights share similar mathematical form (apart from the spin weight parameter $s$) \cite{teukolsky1973perturbations} and the zero-damping quasinormal modes for fields with different spin weights are qualitatively similar \cite{Yang:2012pj,Yang:2013uba}, we expect the result of this analysis also qualitatively applies for fields with other spin weights. It is however important to perform a  study for general fields in more complete settings, including those with both gravitational and additional field perturbations. Nevertheless, the goal of this analysis is to further motivate the conjecture that extremal black holes are critical solutions of a new type of gravitaitonal critical collapse.

Notice that although the full evolution of a gravitational critical collapse should be nonlinear, with fine-tuned initial data there is a stage of evolution where the spacetime/fields can be described by the critical solution and its perturbations, as the initial data is weakly perturbed from the one that gives rise to the exact critical solution. More detailed discussion can be found in section 2 of \cite{Gundlach:2007gc}.

Let us now consider generic extremal black hole spacetimes (with or without assuming GR) with $SO(2,1)\times U(1)$ symmetry at the critical points of associated gravitational collapses. The background metric can be written as \cite{Astefanesei:2006dd}
\begin{align}\label{eq:bg1}
ds^2 =& M^2 \left [ v_1(\theta) \left ( -r^2 dt^2+dr^2/r^2 
  +\beta^2 d\theta^2\right ) \right . \nonumber \\
&  \left . +\beta^2 v_2(\theta) (d \phi +\alpha r dt)^2\right ]
\end{align}
where $v_1, v_2$ are positive functions of $\theta$ and $\alpha, \beta$ are constants. In particular, the NEK (Near Horizon Extremal Kerr) is achieved by setting  
\begin{align}
\alpha =1, \quad \beta=1, \quad v_1 =1+u^2, \quad v_2 =4\frac{1-u^2}{1+u^2}
\end{align}
with $u :=\cos \theta$. For simplicity we consider scalar field evolution on top of this background, mimicking the perturbing fields away from the critical configuration. $\Box \psi =0$ implies that (with $\psi = R(r) S(\theta) e^{i \omega t -i m \phi}$)
\begin{align}
& 2m^2  \left [ \alpha^2 \beta^2 -v_1(\theta)/v_2(\theta)\right ] +4 m \alpha \beta^2 \omega/r  +2 \beta^2 \omega^2/r^2 \nonumber \\
&+ 4r \beta^2\frac{R'}{R} +[\log v_1(\theta) v_2(\theta)]'\frac{S'}{S} +2 r^2 \beta^2\frac{R''}{R} + 2\frac{S''}{S}=0\,,
\end{align}
which leads to two separable equations:
\begin{align}
& 2 S''+ [\log v_1(\theta) v_2(\theta)]'S' +2m^2  \left [ \alpha^2 \beta^2 -v_1(\theta)/v_2(\theta)\right ] S \nonumber \\
& =K \beta^2 S\,,\nonumber \\
& R''+\frac{2 R'}{r} +\frac{1}{2 r^2}\left [ K +4 m \alpha  \omega/r  +2  \omega^2/r^2 \right ] R=0\,.
\end{align}

The angular eigenvalue $K$ can be tuned by choosing different $v_1, v_2,\alpha,\beta, m$. Denoting $\delta^2=1/4-K/2$, in the case of Kerr, it has been shown that $\delta^2$ can be positive or negative, depending on the mode indices $l$ and $m$ of the angular eigenvalue problem \cite{Yang:2012pj,Yang:2013uba}. For generic extremal black holes, the flexibility in $v_1, v_2,\alpha$ and $\beta$ should allow tunable sign of $\delta^2$ even for the same $m$. The general solution of the radial wave equation is \cite{Gralla:2018xzo}
\begin{align}
R_{\rm in} =W_{i m \alpha, i\delta }(-2 i \omega/r), \quad R_{\rm out} = W_{-i m \alpha, -i\delta }(2 i \omega/r)
\end{align}
or ($h_{\pm} =1/2\pm i \delta $)
\begin{align}
R_{\pm} =(-2 i \omega)^{-h_{\pm}} M_{i m \alpha, \pm i \delta} (-2 i \omega/r)\,,
\end{align}
as expressed by Whittaker functions \cite{olver2016nist}. A generic homogeneous solution is a linear combination of any two of these solutions. When $r \rightarrow 0$, we have
\begin{align}
R_{\rm in} \rightarrow e^{i \omega/r} (-2 i \omega/r)^{i m \alpha },\quad R_{\rm out} \rightarrow e^{-i \omega/r} (2 i \omega/r)^{-i m \alpha }
\end{align}
and ($r \rightarrow \infty$)
\begin{align}
R_{\rm \pm} \rightarrow (-2 i \omega/r)^{h_{\pm}}\,. 
\end{align}

For a physical black hole spacetime, the boundary condition should be ingoing at the horizon, so $R_{\rm in}$ should be used. The boundary condition at infinity may be a mixture of ingoing and outgoing waves, depending on the source at infinity (or outside this ``NHEK" region ). If the amplitude ratio between the $\pm$ pieces is $\mathcal{N}$,  the Green's function $g^{\rm mix}_{B\partial}$ can be obtained using similar approach in \cite{Gralla:2018xzo}, which has the time-domain form as
(assuming $\mathcal{E}/\mathcal{N}<e^{-\pi \delta}$)
\begin{widetext}
\begin{align}\label{eq:green}
& G_{\rm  in,mix}(t,r,r'\rightarrow \infty) \propto (\mathcal{N} {r'}^{-h_+}+{r'}^{-h_-}) r^{h_+} \left ( \frac{tr-1}{2}\right ) \Theta (tr-1) \nonumber \\
& \times \sum^\infty_{n=0} \left( \frac{\mathcal{E}}{\mathcal{N}}\right )^n \left ( \frac{tr-1}{2}\right )^{-2 i n \delta} r^{2 i n \delta }
\times  {}_2\tilde{F}_1(h_+-i m\alpha,h_--i m \alpha, h_--i m \alpha +(1-2h_+)n,-(tr-1)/2)\,,
\end{align} 
\end{widetext}
where ${}_2\tilde{F}_1(a,b,c;z):={}_2F_1(a,b,a;z)/\Gamma(c)$, $\mathcal{E} = (h_+-h_-)\Gamma(1-2h_+)\Gamma(h_+-i m \alpha)/[\Gamma(2h_+)\Gamma(h_--i m \alpha)]$,  and ${}_2F_1$ is the hypergeometric function. If the inner boundary condition is taken to be outgoing, the Green's function will contain terms as $r^{ (h_--2 i n \delta)}$. If the boundary condition at $r\rightarrow \infty$ is either ingoing or outgoing, the summation is truncated with only $n=0$ term.
In general we consider the signal which scales as
\begin{align} \label{eq: field solution}
\psi \sim C \sum^\infty_{n=0} r^{ (h_++2 i n \delta)} f_n(tr)  \,,
\end{align}
where $tr$ is the conformally invariant coordinate. This field solution suggests a critical exponent of $1/2$ and the DSS period being $\Delta =2 \pi/\delta$  (as encoded in  $h_{\pm}$) if $\delta$ is real.
When $\delta$ is imaginary, only one of the terms  dominates in the $r \rightarrow 0$ limit. So the whole process becomes CSS when $\delta$ is imaginary. Notice that the sign of $\delta^2$ is tunable for different extremal spacetimes, which all satisfy the spacetime symmetry  $SO(2,1) \times U(1)$, so that changing theories while keeping the same symmetry allows  the DSS period to vary and the transition between DSS and CSS signatures to happen, while the critical component $1/2$ is unaffected (see Fig. \ref{fig: phase diagram}). In addition, it is known that  perturbations around extremal Kerr the corresponding $\delta$ depend on the spin weight of the field \cite{Yang:2013uba}. In this sense the DSS period is non-universal. Similar signatures of the transition between DSS and CSS solutions have been reported in the collapse experiment of Einstein-SU$(2)$ sigma model \cite{Lechner:2001ng}, which also displays diverging $\Delta$ near the transition.

The Choptuik-type critical points are the marginal naked singularity spacetimes satisfying $g_{\mu\nu} =r^2 \tilde{g}_{\mu\nu}(\tilde{x}_i)$ in the CSS regime, with scaling index being two, $\tilde{x}_i :=x_i/r$ being the scale-invariant coordinates and 
\begin{align}
 g_{\mu\nu} =r^2 \tilde{g}_{\mu\nu}(\tilde{x}_i, \log r)\,, \,\, \tilde{g}_{\mu\nu}(\tilde{x}_i, \log r+\Delta) =\tilde{g}_{\mu\nu}(\tilde{x}_i,\log r)
\end{align}
in the DSS regime with period $\Delta$. The function $\tilde{g}$ is non-universal as it varies in different theories.
The extremal-black-hole-type critical points in Eq.~\eqref{eq:bg1}, on the other hand, satisfy the conformal symmetry with scaling index zero. In analogy with the Choptuik-type critical points, it is reasonable to speculate that certain GCCs give rise to extremal black hole critical points with discrete conformal symmetry: $g_{\mu\nu}(t,r,\theta,\phi) = g_{\mu\nu}(t/\lambda, r\lambda,\theta, \phi)$ only for $\lambda=e^{\pm n \Delta}, n=0,1,2,3...$. It is theoretically interesting to search for these DSS extremal black hole solutions from numerical experiments.

\vspace{0.2cm}
\subsection{RG perspective} \label{sec:rgp}

As discussed, depending on the initial data, Eq. \eqref{eq: field solution} describes a stage of evolution as described by the critical solution and its perturbations, which should eventually lead to a Kerr BH, a flat spacetime, or the extremal Kerr BH, \ie the critical point. Fixing $tr$, Eq. \eqref{eq: field solution} has a characteristic length scale, denoted by $L$. According to the usual phenomenology of continuous phase transitions \cite{Cardy1996, Sachdev2011}, whenever the system is at or near the criticality, the physics at length scales $r_0\ll r\ll L$ qualitatively agrees with the criticality, with $r_0\ll L$ certain short-distance cutoff. This picture enables us to reformulate our gravitational collapse via RG.  We will see that the transition with CSS can be described by a stable \underline{UV} fixed point. As $\delta^2$ increases, this stable fixed point collides with unstable ones, leaving behind an RG limit cycle corresponding to the transition with DSS.

We start with the regime where $\delta^2<0$ and the system exhibits emergent CSS at criticality. We view each term in Eq. \eqref{eq: field solution} as an RG fixed point. In the regime $r_0\ll r\ll L$ with $tr$ fixed, the largest term in Eq. \eqref{eq: field solution} becomes the stable UV fixed point describing the physics at criticality, while the presence of other terms is effectively a relevant perturbation. Motivated by \cite{Kaplan2009}, we introduce a running coupling, a central notion in RG, as follows. We modify the wave equation within $r\leqslant r_0$ to be (for variables $r, \xi=rt,\theta,\phi$)
\begin{align}
\frac{\partial^2 \psi}{\partial r^2} +\frac{2}{r} \frac{\partial \psi}{\partial r} +\frac{\lambda}{r^2_0}\psi =0
\end{align}
where the partial derivative here assumes fixed $\xi,\theta,\phi$, and $\lambda$ is a coupling constant. For the relevant $r$-dependent part,
\begin{align}\label{eq:mod}
R''+\frac{2R'}{r} +\frac{\lambda}{r^2_0} R=0\,.
\end{align}

The solution of this equation, with the requirement of regularity at the origin, is
\begin{align}
R \sim \left ( \frac{r}{r_0} \right )^{-1/2} J_{1/2}(\sqrt{\lambda} r/r_0)\,.
\end{align}
Matching this solution with the exterior solution at $r=r_0$ yields
\begin{align}
\frac{r_0 R'(r_0)}{R(r_0)} =\frac{\sqrt{\lambda} J_{3/2}(\sqrt{\lambda})}{J_{1/2}(\sqrt{\lambda})} \equiv \tilde \lambda\,.
\end{align}
Changing the cutoff $r_0$ while demanding the solution at scales $r\gg r_0$ to be independent of $r_0$ requires $\tilde\lambda$ to change with $r_0$, which defines the RG flow of $\tilde\lambda$ as $r_0$ varies.

More concretely, once the boundary condition in the UV is modified, besides the terms in Eq.~\eqref{eq: field solution}, terms that scale as $r^{h_--2i n\delta }$ should also appear, which are also viewed as fixed points. Consider the fixed point corresponding to $\psi_n^{(\pm)}\sim  r^{h_\pm\pm2in\delta}f^{(\pm)}_n(\xi)$. Evaluating $r\psi'/\psi$ at $r=r_0$, we get $\tilde\lambda=\tilde\lambda_\pm^{(n)}$, with
\beq \label{eq: CSS fixed points}
\tilde\lambda_\pm^{(n)}\equiv h_\pm\pm 2in\delta
\eeq
So $\tilde\lambda$ is independent of $r_0$, corroborating our identification of each $\psi_n^{(\pm)}$ with a fixed point. Now consider the field configuration composed with two such fixed points, \eg $\psi\sim c\psi_{n_1}^{(\pm)}+d\psi_{n_2}^{(\pm)}$, with $c$ and $d$ constants. Evaluating $r\psi'/\psi$ at $r=r_0$, we get
\beq \label{eq: pre beta function}
\frac{cf_{n_1}^{(\pm)}(\xi)}{df_{n_2}^{(\pm)}(\xi)}=-r_0^{\tilde\lambda_\pm^{(n_2)}-\tilde\lambda_\pm^{(n_1)}}\cdot\frac{\tilde\lambda-\tilde\lambda_\pm^{(n_2)}}{\tilde\lambda-\tilde\lambda_\pm^{(n_1)}}
\eeq
Define the RG time by $\ell\equiv-\log r_0$, and the beta function of $\tilde\lambda$ by $\beta(\tilde\lambda)\equiv\frac{d\tilde\lambda}{d\ell}$. Differentiating both sides of Eq. \eqref{eq: pre beta function} with respect to $\ell$ yields
\beq
\beta(\tilde\lambda)=\left(\tilde\lambda-\tilde\lambda_\pm^{(n_1)}\right)\left(\tilde\lambda-\tilde\lambda_\pm^{(n_2)}\right)
\eeq
which means that the fixed point with a smaller (larger) $\tilde\lambda$ is more stable (unstable) in the UV. This is expected from our RG perspective, because in the UV the field is dominated by the largest term, and $\tilde\lambda$ should flow to the value characterizing the fixed point corresponding to this term.

As $\delta^2$ is tuned from a negative value to zero, all these fixed points collide at $\tilde\lambda=1/2$. Further tuning $\delta^2$ to be positive pushes the system into the DSS regime, where Eq. \eqref{eq: CSS fixed points} becomes complex, and all terms in Eq. \eqref{eq: field solution} are oscillating and equally important. So we should consider the following field configuration as describing the critical point
\begin{align}\label{eq:gen}
\psi \sim c \sum_n r^{h_++ 2i n \delta } f_n(\xi) +d \sum_n r^{h_--2i n \delta} g_n(\xi)
\end{align}
with $c,d$ being constants and $f_n, g_n$ determined by the Green's function.
Evaluating $r \psi'/\psi$ at $r=r_0$ gives
\begin{widetext}
\begin{align}\label{eq:inb}
\tilde\lambda = \frac{c/d \sum_n (h_++ 2i n \delta) r_0^{h_++ 2i n \delta } f_n(\xi) + \sum_n (h_--2i n \delta) r_0^{h_--2i n \delta} g_n(\xi)}{c/d \sum_n r_0^{h_++ 2i n \delta } f_n(\xi) + \sum_n r_0^{h_--2i n \delta} g_n(\xi)}
\end{align}
\end{widetext}
In this case, $\tilde\lambda$ does run as $r_0$ changes. However, it is observed that $\tilde\lambda$ returns to itself as $r_0\rightarrow r_0e^{\frac{2\pi}{\delta}}$, which means that $\tilde\lambda$ undergoes an RG limit cycle, with period $2\pi/\delta$ (measured in $\ell=-\log r_0$).

It is reasonable to expect that such arguments carry through for the critical spacetimes as well, in addition to their perturbations. We notice that the key to this process is to realize a solution with a form similar to  Eq.~\eqref{eq:gen}, i.e., periodicity in the scaling dimension. In the case of Choptuik spherical collapse, the scale-invariant coordinate is $\xi =t/r$
and ${\rm Re}(h_\pm)$ depends on the nature of fields considered (two for the metric and zero for the collapsing scalar field). If the wave equations for the metric quantities and the scalar field are modified in a similar way as Eq.~\eqref{eq:mod} within certain cutoff radius $r_0$ and arbitrary $\xi$, it is expected that the critical solution depends on $r_0$ and the inner ($r\leqslant r_0$) physics assumed. As the outer (critical) solution is periodic in the $\log r$ direction, it should return to the original state if $r\rightarrow r e^{2 \pi/\delta}$, i.e., a RG limit cycle with $\Delta=2\pi/\delta$. The direct implementation and demonstration for Choptuik-type GCC is left for future work.

\vspace{0.2cm}
\subsection{Gravitational Efimov Effect.}
In systems with RG limit cycles, after a UV cutoff $\Lambda_{\rm UV}$ is imposed, a series of IR scales generically emerges with $\Lambda_{\rm IR} \sim \Lambda_{\rm UV} e^{-n\Delta}$ \cite{Kaplan2009}. In particular, the ``Efimov states" may appear forming a geometric spectrum, as associated with the IR scales. In gravitational systems, we now discuss a similar phenomenon which is  present for extremal black holes. In particular, we shall show that, if a UV cutoff is placed at $x_0 = r/M-1 \ll 1$ for an extremal Kerr black hole, there is a family of quasinormal modes with frequencies forming a geometric series:
\begin{align}
\omega_n\propto 1/\sqrt{x_0}, \quad \omega_n =\omega_0 e^{n \pi/(2\delta)}\,,
\end{align}
with $n=0,1,2,3 ..$. The Choptuik-type critical points should have emergent length scales as $r_0 e^{n \Delta}$ if the UV cutoff is placed at $r_0$, because of the log periodicity of the field. It is possible to form (quasi)-bound states with frequencies in geometric series as well. 

In Efimov systems, the cutoff at the UV scale naturally introduces an IR scale and there is an infinite ladder of bound states with energies being associated with the IR scale and given by a geometric series.
Let us search for similar signatures in extremal black holes, e.g. extremal Kerr. We adopt the Boyer-Lindquist coordinate and construct a UV cutoff boundary condition $ x \psi'/\psi =\gamma$ at $r=r_0$ ($x\equiv r-1, x_0=r_0-1\ll 1$). Here for simplicity we have set the black hole mass $M=1$. If we view an extremal Kerr black hole as the limit of near-extremal Kerr with $a \rightarrow 1$, 
\begin{align}
\partial_t/\kappa \rightarrow T\partial_T-R\partial_R
\end{align}
where $T,R$ are the time and radial coordinate in the NHEK spacetime (which is labeled as $t,r$ in Eq.~$2$ in the main text) and $\kappa \equiv \sqrt{1-a^2}$. In other words, $\partial_t$ is the killing vector field that corresponds to the conformal invariant coordinate $\xi$. To be compatible with Eq.~$13$ in the main text, where $\partial_r$ is respect to fixed $\xi$, a mode analysis should consider the Fourier transform of $t$ instead of $T$.
Such equation is just the Teukolsky equation \cite{Yang:2013uba} (which is also true for near-extremal black holes with $x \gg \kappa$):
\begin{align}\label{eq:tf}
x^2 R''+2xR'+\left[\omega^2(x+2)^2 -\lambda\right ] R=0,
\end{align}
where $\lambda = A_{\ell m \omega}+\omega^2-2 m \omega$ with $ A_{\ell m \omega}$ being the eigenvalue of the angular Teukolsky equation. The homogeneous solutions of this equation are
\begin{align}\label{eq:far}
R =&  A e^{-i \omega x} x^{-1/2+i\delta} \times {}_{1}F_1(1/2+i \delta+2i \omega, 1+2 i \delta, 2 i \omega x) \nonumber \\ 
&+ B(\delta \rightarrow -\delta) 
\end{align}
where ${}_{1}F_1$ is the confluent hypergeometric function and we only consider the DSS regime with $\delta >0$.
The outgoing boundary condition at $r \rightarrow \infty$ suggests that
\begin{align}\label{eq:farratio}
\frac{A}{B} = e^{\pi \delta + 2 i \delta \log 2\omega} \frac{\Gamma(-2i\delta) \Gamma(1/2+i \delta -2 i \omega)}{\Gamma(2 i \delta) \Gamma(1/2-i \delta -2 i \omega)}\,.
\end{align}
We shall write the above expression as
\begin{align}\label{eq:farratio2}
&\frac{A}{B} \approx e^{\pi \delta + 4 i \delta \log 2\omega} \left [e^{ -2 i \delta \log 2\omega}\frac{\Gamma(-2i\delta) \Gamma(1/2+i \delta -2 i \omega)}{\Gamma(2 i \delta) \Gamma(1/2-i \delta -2 i \omega)} \right ] 
\nonumber \\
&:=e^{\pi \delta + 4 i \delta \log 2\omega} f_\delta\,,
\end{align}
as for $|\omega|\ge 1$, the term in the square bracket turns out to be approximately independent of $\omega$ as we change the magnitude of $\omega$. Now consider the boundary condition at $r_0$, we have
\begin{align}
\frac{A (-1/2+i \delta) x^{-1/2+i \delta}_0+B (-1/2-i \delta) x^{-1/2-i \delta}_0}{A  x^{-1/2+i \delta}_0+B  x^{-1/2-i \delta}_0}=\gamma
\end{align}
or
\begin{align}
\frac{A/B (-1/2+i \delta) x^{2i \delta}_0+ (-1/2-i \delta) }{A/B  x^{2i \delta}_0+  1}=\gamma\,.
\end{align}

It is straighforward to see that, $\omega \propto 1/\sqrt{x_0}$ and if $\omega_0$ is a solution of this equation, then $\omega_n =\omega_0 e^{n\pi/(2\delta)}$ with $n=0,1,2,..$ should also be a solution. Therefore we see the 
quasinormal mode frequencies for the spacetime with a UV cutoff also form a geometric series. 
It is interesting to search for similar signatures in Choptiok-type critical spacetime as well.

\vspace{0.2cm}
\section{RG in non-gravitational systems.} RG has been studied in various non-gravitational contexts. Fixed points are more familiar, and it is well known that they lead to CSS, similar to our GCC with $\delta^2<0$ \cite{Peskin1995, Cardy1996, Wen2004, Sachdev2011}. Limit cycles have also been studied in quantum field theory \cite{Wilson1971, Bernard2001, LeClair2003a,LeClair2003, Jepsen2020}, statistical physics \cite{Huse1991, Veytsman1993}, quantum few-body systems \cite{Bedaque1998, Glazek2002, Kaplan2009} and quantum many-body systems \cite{Hartnoll2015, Yerzhakov2020}. See Ref. \cite{Bulycheva2014} for a review. Limit cycles often manifest themselves as some DSS behavior, such as Efimov effect and log-periodic behavior at or near continuous phase transitions.

To highlight the similarity between our GCC with $\delta^2>0$ and continuous phase transitions in non-gravitational systems described by RG limit cycles, here we first summarize phenomena related to RG limit cycles and DSS. The detailed derivation is presented in the later part of this section. 

Suppose a continuous phase transition is accessed by tuning a parameter $p$ to the critical point at $p=0$, and the critical theory has a parameter $\theta$ that undergoes an RG limit cycle. The correlation functions exhibit DSS at criticality. For example, the two-point correlation function $G(k; p=0, \theta)$ at momentum $k$ satisfies that $G(sk; p=0, \alpha)=s^{c_1}G(k; p=0, \theta)$ only for specific choices of $s$ and a constant $c_1$, which further implies that $G$ can be written as $G(k; p=0, \theta)=k^{c_1}\tilde G_\theta(\log k)$, where $\tilde G_\theta$ is a periodic function that depends on $\theta$. Note this DSS behaviour of correlation function is similar to the DSS Green's function described by Eq.~\eqref{eq:green}. Away from the critical point, the correlation length $L$ diverges as $L\sim p^{-c_2}\tilde L_\theta(\log p)$, and the order parameter (if any) $M\sim p^{c_3}M_\theta(\log p)$, where $c_{2, 3}$ are also universal constants, and $L_\theta$ and $M_\theta$ are $\theta$-dependent periodic functions.

The above phenomena show remarkable similarity with GCC, which further supports our RG perspective on the latter.


We now derive of the consequences of the presence of a coupling that undergoes an RG limit cycle on the critical behaviors in a continuous phase transition. We will see, from Eqs. \eqref{eqapp: DSS of correlation function} and \eqref{eqapp: log-periodic correlation function}, that the correlation functions of the system right at the critical point shows discrete self similarity (DSS), in contrast to the usual continuous self similarity (CSS) that shows up at critical points without any coupling undergoing an RG limit cycle. Moreover, according to Eqs. \eqref{eqapp: correlation length} and \eqref{eqapp: order parameter}, the singularities of physical quantities that appear when the system approaches the critical point is also not given by the usual power law of the deviation from the critical point, but such a power law multiplied by a log-periodic function of the deviation, \ie a periodic function of the logarithm of the deviation. The period of this log-periodic function is determined by the period of the limit cycle, and its other aspects depend on the precise value of the parameter that undergoes an RG limit cycle.

For concreteness, we consider a continuous phase transition that can be described by a renormalizable field theory in $D$ dimensional Euclidean spacetime, which has the full Euclidean symmetry (\ie translation and rotation symmetries). It is straightforward to generalize the consideration here to other field theories, which may break the Euclidean symmetry.

We assume that the transition is accessed by tuning a parameter $p$ to the critical point at $p=0$. There may or may not be a local order parameter associated with this transition, but if there is (\eg the transition is related to a spontaneous symmetry breaking), we denote this order parameter by $m$, which is coupled to a source denoted by $h$. Besides the couplings $p$ and $h$, there is another single coupling constant $\theta$ that undergoes an RG limit cycle. The beta functions of these couplings near the critical point take the form
\beq
\begin{split}
&\beta(p)\equiv-\mu\frac{dp}{d\mu}=\Delta_pp\\
&\beta(h)\equiv-\mu\frac{dh}{d\mu}=\Delta_hh\\
&\beta(\theta)\equiv-\mu\frac{d\theta}{d\mu}
\end{split}
\eeq
where $\mu$ is the renormalization scale, and the dimensionless constants $\Delta_p$ and $\Delta_h$ are the scaling dimensions of $p$ and $h$, respectively. We do not need to specify the details of $\beta(\theta)$, except that it induces a limit cycle of $\theta$, \ie the running $\theta$ is a periodic function of $\log\mu$. In the above, we have ignored the corrections to the beta function of one coupling due to the other couplings; in particular, we have assumed that the beta function of $\theta$ only depends on $\theta$, but not on $p$ or $h$. This should be examined for each specific field theory. We expect that these corrections are often small as long as the system is sufficiently close to the transition, where $p=0$ and $h=0$. We have also assumed that all other couplings are irrelevant, and their amplitudes can be ignored when the system is sufficiently close to the critical point.

\subsection{Discrete self similarity in the correlation functions at the transition}

We first demonstrate that the correlation functions of the system at the transition, \ie $p=0$ and $h=0$, exhibits DSS. To show this, we note that the correlation functions generically satisfy a Callan-Symanzik equation \cite{Peskin1995}. To be concrete, consider a two-point correlation function in momentum space, $G(k; \theta, \mu)$, where $k$ is a momentum. The Callan-Symanzik equation takes the form
\beq
\left[k\frac{\partial}{\partial k}+\beta(\theta)\frac{\partial}{\partial\theta}-\Delta(\theta)\right]G(k; \theta, \mu)=0
\eeq
where the function $\Delta(\theta)$ is determined by the dynamics of the field theory, just like $\beta(\theta)$. From this equation, we can already see the usual result applicable to systems described by an RG fixed point. Suppose the fixed point is at $\theta=\theta_*$, such that $\beta(\theta_*)=0$. Then $G(k; \theta_*, \mu)\propto |k|^{\Delta(\theta_*)}$. So the correlation function displays CSS, \ie $G(sk; \theta_*, \mu)=s^{\Delta(\theta_*)}G(k; \theta_*, \mu)$ for {\it any} $s$, $k$ and $\mu$.

Our system is not described by an RG fixed point, but a limit cycle. To extract the property of the correlation function in this case, we look at the solution to this Calllan-Symanzik equation \cite{Peskin1995} (see Eqs. (12.72) and (12.73) therein):
\beq \label{eqapp: correlation function solution}
G(k; \theta, \mu)=\hat G(\bar\theta(k; \theta))\exp\left[\int_{k'=\mu}^{k'=k}d\log\frac{k'}{\mu}\Delta(\bar\theta(k'; \theta))\right] \nonumber \\
\eeq
where $\bar\theta(k; \theta)$ satisfies
\beq
\frac{d}{d\log\frac{k}{\mu}}\bar\theta(k; \theta)=-\beta(\bar\theta),
\quad
\bar\theta(\mu; \theta)=\theta
\eeq
and $\hat G$ is a function that sets the initial condition of the Callan-Symanzik equation, which is also determined by the dynamics of the field theory.

That $\theta$ undergoes an RG limit cycle means that there exists a {\it specific} constant $s$ such that $\bar\theta(sk; \theta)=\bar\theta(k; \theta)$ for any $k$ and $\theta$. Then Eq. \eqref{eqapp: correlation function solution} implies that
\beq
\begin{split}
G(sk; \theta, \mu)
&=\hat G(\bar\theta(sk; \theta))\exp\left[\int_{k'=\mu}^{k'=sk}d\log\frac{k'}{\mu}\Delta(\bar\theta(k'; \theta))\right]\\
&=G(k; \theta, \mu)\exp\left[\int_{k'=k}^{k'=sk}d\log\frac{k'}{\mu}\Delta(\bar\theta(k'; \theta))\right]
\end{split}
\eeq
The limit cycle behavior of $\bar\theta$ further implies that the last factor depends only on $s$ and the precise structure of the function $\Delta(\bar\theta)$, but not on the value of $k/\mu$ or $\theta$. Denoting the last factor by $s^{\bar\Delta}$ where $\bar\Delta$ is a dimensionless constant, we get the DSS behavior of the correlation function
\beq \label{eqapp: DSS of correlation function}
G(sk; \theta, \mu)=s^{\bar\Delta}G(k; \theta, \mu)
\eeq
where $s$ is a specific value, while $k$, $\theta$ and $\mu$ can be arbitrary.

It may be useful to write the correlation function as $G(k; \theta, \mu)\equiv k^{\bar\Delta}\hat G(k; \theta, \mu)$. Then Eq. \eqref{eqapp: DSS of correlation function} implies that $\hat G(sk; \theta, \mu)=\hat G(k; \theta, \mu)$. Now define $\tilde G(k; \theta, \mu)=\hat G\left({\color{blue}\mu}e^{k/\mu}; \theta, \mu\right)$, then $\tilde G(k+\log s; \theta, \mu)=\tilde G(k; \theta, \mu)$ and
\beq \label{eqapp: log-periodic correlation function}
G(k; \theta, \mu)=k^{\bar\Delta}\tilde G\left(\log\frac{k}{\mu}; \theta, \mu\right)
\eeq
That is, such a correlation function with DSS can be written as a power law multiplied by a log-periodic function with a period. Note that $\tilde G$ still depends on $\theta$ even at the transition.

\subsection{Singularities of physical quantities}

In a continuous phase transition described by an RG fixed point, various physical quantities show power-law singularity as the system approaches the critical point. For example, the correlation length diverges as a power law of the deviation from the critical point. Now we examine how these singularities are modified in the presence of a coupling that undergoes an RG limit cycle.

Let us start with the correlation length $\xi$. To this end, it suffices to consider the case $h=0$. In general, $\xi$ is a function of $p$ and $\theta$, which in turn depend on the RG parameter $\ell\equiv-\log\frac{\mu}{\mu_0}$, where $\mu_0$ is an arbitrary momentum scale to make the argument of this logarithm dimensionless. By definition, the correlation length satisfies that
\beq
\frac{d\xi}{d\ell}=-\xi
\eeq
so
\beq
\frac{\partial\xi}{\partial p}\frac{dp}{d\ell}+\frac{\partial\xi}{\partial\theta}\frac{d\theta}{d\ell}=\Delta_pp\frac{\partial\xi}{\partial p}+\beta(\theta)\frac{\partial\xi}{\partial\theta}=-\xi
\eeq
The solution of this equation is
\beq \label{eqapp: correlation length}
\xi(p, \theta)=\left(\frac{p_0}{p}\right)^{\frac{1}{\Delta_p}}\cdot g\left(\bar\theta(p, \theta)\right)
\eeq
where $p_0$ and $g$ are determined by the boundary condition of the above equation, and $\bar\theta(p, \theta)$ satisfies 
\beq \label{eqapp: theta bar}
\frac{d\bar\theta(p, \theta)}{d\log p}=-\frac{\beta(\bar\theta)}{\Delta_p},
\quad
\bar\theta(p_0, \theta)=\theta
\eeq

For the case of an RG fixed point where the only symmetric relevant coupling is $p$, the function $g$ can be viewed as a constant since there is no notion of $\theta$, and we see that the correlation length diverges as $\xi\sim p^{-\frac{1}{\Delta_p}}$, which is the standard result \cite{Cardy1996}  (see Eq. (3.50) therein). In our case, $\theta$ undergoes an RG limit cycle, so $\xi\sim p^{-\frac{1}{\Delta_p}}\tilde g_\theta(\log p)$, where $\tilde g_\theta$ is a periodic function of $\log p$ and it depends on $\theta$.

Next, we turn to the order parameter. To this end, we need to consider a nonzero $h$ and the free energy density $f(p, h, \theta)$. The free energy density is simply the generating functional of the field theory with an external source that is uniform in spacetime. It satisfies that
\beq
\frac{df}{d\ell}=Df
\eeq
where $D$ is the spacetime dimension. So
\begin{align}
&\frac{\partial f}{\partial p}\frac{dp}{d\ell}+\frac{\partial f}{\partial h}\frac{dh}{d\ell}+\frac{\partial f}{\partial\theta}\frac{d\theta}{d\ell}=\Delta_p p\frac{\partial f}{\partial p}+\Delta_h h \frac{\partial f}{\partial h}+\beta(\theta)\frac{\partial f}{\partial\theta} \nonumber \\
&=Df
\end{align}
The solution of this equation is
\beq \label{eqapp: free energy density}
f(t, h, \theta)=\left(\frac{p}{p_0}\right)^{\frac{D}{\Delta_p}}\cdot\hat f(\bar h(p, h), \bar\theta(p, \theta))
\eeq
where $p_0$ and $\hat f$ are again determined by the boundary condition of the above equation, $\bar\theta$ still satisfies Eq. \eqref{eqapp: theta bar}, and $\bar h(p, h)=h\cdot\left(p_0/p\right)^{\Delta_h/\Delta_p}$.

For the usual case of an RG fixed point, we can drop $\theta$ in Eq. \eqref{eqapp: free energy density}, which then simplifies as $f(p, h)=\left(\frac{p}{p_0}\right)^{\frac{D}{\Delta_p}}\cdot\hat f(h\cdot\left(p_0/p\right)^{\Delta_h/\Delta_p})$. The order parameter is given by the standard result $m=\left.\frac{\partial f}{\partial h}\right|_{h=0}\sim p^{\frac{D-\Delta_h}{\Delta_p}}$ \cite{Cardy1996} (see Eq. (3.31) therein). In our case, the order parameter
\beq \label{eqapp: order parameter}
m=\left.\frac{\partial f}{\partial h}\right|_{h=0}=\left(\frac{p}{p_0}\right)^{\frac{D-\Delta_h}{\Delta_p}}\cdot\frac{\partial\hat f(0, \bar\theta(p, \theta))}{\partial \bar h}
\eeq
Because $\theta$ undergoes an RG limit cycle, the second factor in the above expression is a periodic function of $\log t$, which further depends on $\theta$.

In summary, in the presence of a coupling undergoing an RG limit cycle, the continuous self similarity commonly seen in a continuous phase transition is broken to discrete self similarity, and it manifests as some log-periodic multiplicative correction to the usual power-law behavior.

\vspace{0.2cm}
\section{Discussion.}
If we treat CSS solutions as a special case of DSS solutions with $\Delta=\infty$, the critical signatures of a GCC should at least include the scaling index of the critical spacetime, the DSS period $\Delta_1$, the critical exponent of the dominant perturbation mode $\gamma$ and the DSS period $\Delta_2$ of the perturbation. There are GCC cases where $\Delta_1=\Delta_2$ (e.g., Choptuik collapse \cite{Choptuik:1992jv} ) and cases where $\Delta_1=\infty, \Delta_2 \neq \infty$ (collapse with extremal black holes). It is unclear whether the case(s) with $(\Delta_1 \neq \Delta_2 )\neq \infty$ is possible in GCCs. Traditionally the universality classes have been used to classify the critical phenomena and distinguish between different kinds of critical systems. It is unclear whether similar concepts can be applied for GCC systems. In particular, as DSS periods are non-universal, do the scaling index and the critical exponent uniquely define a universality class, or if other observables (such as the correlation function) are also required to compare the critical behaviour in two distinct systems? Such understanding will be crucial if the comparison is extended to be made between gravitational and non-gravitational systems.

The Choptuik-type critical points have the scaling index being two and the extremal black holes have the scaling index being zero.  It is natural to ask, whether they are the only viable critical points for GCCs in four dimensions? The answer is possibly true because one option of the final state: the black hole fixed point, can all be described by the mass, spin and charge according to
the no hair theorem. The marginal black hole formation may either correspond to a zero-mass naked singularity or an extremal black hole. As the Efimov systems discussed in \cite{Bulycheva2014} often have symmetry being ${\rm SO}(2,1)$, i.e., similar to the extreme black hole scenario, it will be interesting to search for other systems with RG limit cycles that have similar signatures as Choptuik-type critical points.

\vspace{0.2cm}
\noindent{\it Acknowledgements.}
HY was supported by the National Science and Engineering Research Council through a Discovery grant.  HY thanks KITP where part of the work was completed. This research was supported in part by the National Science Foundation under Grant No. NSF PHY-1748958. This research was also supported in part by Perimeter Institute for Theoretical Physics. Research at Perimeter Institute is supported by the Government of Canada through the Department of Innovation, Science and Economic Development Canada and by the Province of Ontario through the Ministry of Research, Innovation and Science.

\bibliography{ref.bib}

\begin{thebibliography}{33}%
\makeatletter
\providecommand \@ifxundefined [1]{%
 \@ifx{#1\undefined}
}%
\providecommand \@ifnum [1]{%
 \ifnum #1\expandafter \@firstoftwo
 \else \expandafter \@secondoftwo
 \fi
}%
\providecommand \@ifx [1]{%
 \ifx #1\expandafter \@firstoftwo
 \else \expandafter \@secondoftwo
 \fi
}%
\providecommand \natexlab [1]{#1}%
\providecommand \enquote  [1]{``#1''}%
\providecommand \bibnamefont  [1]{#1}%
\providecommand \bibfnamefont [1]{#1}%
\providecommand \citenamefont [1]{#1}%
\providecommand \href@noop [0]{\@secondoftwo}%
\providecommand \href [0]{\begingroup \@sanitize@url \@href}%
\providecommand \@href[1]{\@@startlink{#1}\@@href}%
\providecommand \@@href[1]{\endgroup#1\@@endlink}%
\providecommand \@sanitize@url [0]{\catcode `\\12\catcode `\$12\catcode
  `\&12\catcode `\#12\catcode `\^12\catcode `\_12\catcode `\%12\relax}%
\providecommand \@@startlink[1]{}%
\providecommand \@@endlink[0]{}%
\providecommand \url  [0]{\begingroup\@sanitize@url \@url }%
\providecommand \@url [1]{\endgroup\@href {#1}{\urlprefix }}%
\providecommand \urlprefix  [0]{URL }%
\providecommand \Eprint [0]{\href }%
\providecommand \doibase [0]{http://dx.doi.org/}%
\providecommand \selectlanguage [0]{\@gobble}%
\providecommand \bibinfo  [0]{\@secondoftwo}%
\providecommand \bibfield  [0]{\@secondoftwo}%
\providecommand \translation [1]{[#1]}%
\providecommand \BibitemOpen [0]{}%
\providecommand \bibitemStop [0]{}%
\providecommand \bibitemNoStop [0]{.\EOS\space}%
\providecommand \EOS [0]{\spacefactor3000\relax}%
\providecommand \BibitemShut  [1]{\csname bibitem#1\endcsname}%
\let\auto@bib@innerbib\@empty
\bibitem [{\citenamefont {Choptuik}(1993)}]{Choptuik:1992jv}%
  \BibitemOpen
  \bibfield  {author} {\bibinfo {author} {\bibfnamefont {Matthew~W.}\
  \bibnamefont {Choptuik}},\ }\bibfield  {title} {\enquote {\bibinfo {title}
  {{Universality and scaling in gravitational collapse of a massless scalar
  field}},}\ }\href {\doibase 10.1103/PhysRevLett.70.9} {\bibfield  {journal}
  {\bibinfo  {journal} {Phys. Rev. Lett.}\ }\textbf {\bibinfo {volume} {70}},\
  \bibinfo {pages} {9--12} (\bibinfo {year} {1993})}\BibitemShut {NoStop}%
\bibitem [{\citenamefont {Gundlach}\ and\ \citenamefont
  {Martin-Garcia}(2007)}]{Gundlach:2007gc}%
  \BibitemOpen
  \bibfield  {author} {\bibinfo {author} {\bibfnamefont {Carsten}\ \bibnamefont
  {Gundlach}}\ and\ \bibinfo {author} {\bibfnamefont {Jose~M.}\ \bibnamefont
  {Martin-Garcia}},\ }\bibfield  {title} {\enquote {\bibinfo {title} {{Critical
  phenomena in gravitational collapse}},}\ }\href {\doibase
  10.12942/lrr-2007-5} {\bibfield  {journal} {\bibinfo  {journal} {Living Rev.
  Rel.}\ }\textbf {\bibinfo {volume} {10}},\ \bibinfo {pages} {5} (\bibinfo
  {year} {2007})},\ \Eprint {http://arxiv.org/abs/0711.4620} {arXiv:0711.4620
  [gr-qc]} \BibitemShut {NoStop}%
\bibitem [{\citenamefont {Gundlach}(1997)}]{Gundlach:1996eg}%
  \BibitemOpen
  \bibfield  {author} {\bibinfo {author} {\bibfnamefont {Carsten}\ \bibnamefont
  {Gundlach}},\ }\bibfield  {title} {\enquote {\bibinfo {title} {{Understanding
  critical collapse of a scalar field}},}\ }\href {\doibase
  10.1103/PhysRevD.55.695} {\bibfield  {journal} {\bibinfo  {journal} {Phys.
  Rev. D}\ }\textbf {\bibinfo {volume} {55}},\ \bibinfo {pages} {695--713}
  (\bibinfo {year} {1997})},\ \Eprint {http://arxiv.org/abs/gr-qc/9604019}
  {arXiv:gr-qc/9604019} \BibitemShut {NoStop}%
\bibitem [{\citenamefont {Martin-Garcia}\ and\ \citenamefont
  {Gundlach}(1999)}]{Martin-Garcia:1998zqj}%
  \BibitemOpen
  \bibfield  {author} {\bibinfo {author} {\bibfnamefont {Jose~M.}\ \bibnamefont
  {Martin-Garcia}}\ and\ \bibinfo {author} {\bibfnamefont {Carsten}\
  \bibnamefont {Gundlach}},\ }\bibfield  {title} {\enquote {\bibinfo {title}
  {{All nonspherical perturbations of the Choptuik space-time decay}},}\ }\href
  {\doibase 10.1103/PhysRevD.59.064031} {\bibfield  {journal} {\bibinfo
  {journal} {Phys. Rev. D}\ }\textbf {\bibinfo {volume} {59}},\ \bibinfo
  {pages} {064031} (\bibinfo {year} {1999})},\ \Eprint
  {http://arxiv.org/abs/gr-qc/9809059} {arXiv:gr-qc/9809059} \BibitemShut
  {NoStop}%
\bibitem [{\citenamefont {Husa}\ \emph {et~al.}(2000)\citenamefont {Husa},
  \citenamefont {Lechner}, \citenamefont {Purrer}, \citenamefont {Thornburg},\
  and\ \citenamefont {Aichelburg}}]{Husa:2000kr}%
  \BibitemOpen
  \bibfield  {author} {\bibinfo {author} {\bibfnamefont {Sascha}\ \bibnamefont
  {Husa}}, \bibinfo {author} {\bibfnamefont {Christiane}\ \bibnamefont
  {Lechner}}, \bibinfo {author} {\bibfnamefont {Michael}\ \bibnamefont
  {Purrer}}, \bibinfo {author} {\bibfnamefont {Jonathan}\ \bibnamefont
  {Thornburg}}, \ and\ \bibinfo {author} {\bibfnamefont {Peter~C.}\
  \bibnamefont {Aichelburg}},\ }\bibfield  {title} {\enquote {\bibinfo {title}
  {{Type II critical collapse of a selfgravitating nonlinear sigma model}},}\
  }\href {\doibase 10.1103/PhysRevD.62.104007} {\bibfield  {journal} {\bibinfo
  {journal} {Phys. Rev. D}\ }\textbf {\bibinfo {volume} {62}},\ \bibinfo
  {pages} {104007} (\bibinfo {year} {2000})},\ \Eprint
  {http://arxiv.org/abs/gr-qc/0002067} {arXiv:gr-qc/0002067} \BibitemShut
  {NoStop}%
\bibitem [{\citenamefont {Koike}\ \emph {et~al.}(1995)\citenamefont {Koike},
  \citenamefont {Hara},\ and\ \citenamefont {Adachi}}]{Koike:1995jm}%
  \BibitemOpen
  \bibfield  {author} {\bibinfo {author} {\bibfnamefont {Tatsuhiko}\
  \bibnamefont {Koike}}, \bibinfo {author} {\bibfnamefont {Takashi}\
  \bibnamefont {Hara}}, \ and\ \bibinfo {author} {\bibfnamefont {Satoshi}\
  \bibnamefont {Adachi}},\ }\bibfield  {title} {\enquote {\bibinfo {title}
  {{Critical behavior in gravitational collapse of radiation fluid: A
  Renormalization group (linear perturbation) analysis}},}\ }\href {\doibase
  10.1103/PhysRevLett.74.5170} {\bibfield  {journal} {\bibinfo  {journal}
  {Phys. Rev. Lett.}\ }\textbf {\bibinfo {volume} {74}},\ \bibinfo {pages}
  {5170--5173} (\bibinfo {year} {1995})},\ \Eprint
  {http://arxiv.org/abs/gr-qc/9503007} {arXiv:gr-qc/9503007} \BibitemShut
  {NoStop}%
\bibitem [{\citenamefont {Kehle}\ and\ \citenamefont
  {Unger}(2024)}]{Kehle:2024vyt}%
  \BibitemOpen
  \bibfield  {author} {\bibinfo {author} {\bibfnamefont {Christoph}\
  \bibnamefont {Kehle}}\ and\ \bibinfo {author} {\bibfnamefont {Ryan}\
  \bibnamefont {Unger}},\ }\bibfield  {title} {\enquote {\bibinfo {title}
  {{Extremal black hole formation as a critical phenomenon}},}\ }\href@noop {}
  {\  (\bibinfo {year} {2024})},\ \Eprint {http://arxiv.org/abs/2402.10190}
  {arXiv:2402.10190 [gr-qc]} \BibitemShut {NoStop}%
\bibitem [{\citenamefont {Sorce}\ and\ \citenamefont
  {Wald}(2017)}]{Sorce:2017dst}%
  \BibitemOpen
  \bibfield  {author} {\bibinfo {author} {\bibfnamefont {Jonathan}\
  \bibnamefont {Sorce}}\ and\ \bibinfo {author} {\bibfnamefont {Robert~M.}\
  \bibnamefont {Wald}},\ }\bibfield  {title} {\enquote {\bibinfo {title}
  {{Gedanken experiments to destroy a black hole. II. Kerr-Newman black holes
  cannot be overcharged or overspun}},}\ }\href {\doibase
  10.1103/PhysRevD.96.104014} {\bibfield  {journal} {\bibinfo  {journal} {Phys.
  Rev. D}\ }\textbf {\bibinfo {volume} {96}},\ \bibinfo {pages} {104014}
  (\bibinfo {year} {2017})},\ \Eprint {http://arxiv.org/abs/1707.05862}
  {arXiv:1707.05862 [gr-qc]} \BibitemShut {NoStop}%
\bibitem [{\citenamefont {Boulware}(1973)}]{Boulware:1973tlq}%
  \BibitemOpen
  \bibfield  {author} {\bibinfo {author} {\bibfnamefont {David~G.}\
  \bibnamefont {Boulware}},\ }\bibfield  {title} {\enquote {\bibinfo {title}
  {{Naked Singularities, Thin Shells, and the Reissner-Nordstr\"om Metric}},}\
  }\href {\doibase 10.1103/PhysRevD.8.2363} {\bibfield  {journal} {\bibinfo
  {journal} {Phys. Rev. D}\ }\textbf {\bibinfo {volume} {8}},\ \bibinfo {pages}
  {2363} (\bibinfo {year} {1973})}\BibitemShut {NoStop}%
\bibitem [{\citenamefont {Teukolsky}(1973)}]{teukolsky1973perturbations}%
  \BibitemOpen
  \bibfield  {author} {\bibinfo {author} {\bibfnamefont {Saul~A}\ \bibnamefont
  {Teukolsky}},\ }\bibfield  {title} {\enquote {\bibinfo {title} {Perturbations
  of a rotating black hole. i. fundamental equations for gravitational,
  electromagnetic, and neutrino-field perturbations},}\ }\href@noop {}
  {\bibfield  {journal} {\bibinfo  {journal} {Astrophysical Journal, Vol. 185,
  pp. 635-648 (1973)}\ }\textbf {\bibinfo {volume} {185}},\ \bibinfo {pages}
  {635--648} (\bibinfo {year} {1973})}\BibitemShut {NoStop}%
\bibitem [{\citenamefont {Yang}\ \emph
  {et~al.}(2013{\natexlab{a}})\citenamefont {Yang}, \citenamefont {Zhang},
  \citenamefont {Zimmerman}, \citenamefont {Nichols}, \citenamefont {Berti},\
  and\ \citenamefont {Chen}}]{Yang:2012pj}%
  \BibitemOpen
  \bibfield  {author} {\bibinfo {author} {\bibfnamefont {Huan}\ \bibnamefont
  {Yang}}, \bibinfo {author} {\bibfnamefont {Fan}\ \bibnamefont {Zhang}},
  \bibinfo {author} {\bibfnamefont {Aaron}\ \bibnamefont {Zimmerman}}, \bibinfo
  {author} {\bibfnamefont {David~A.}\ \bibnamefont {Nichols}}, \bibinfo
  {author} {\bibfnamefont {Emanuele}\ \bibnamefont {Berti}}, \ and\ \bibinfo
  {author} {\bibfnamefont {Yanbei}\ \bibnamefont {Chen}},\ }\bibfield  {title}
  {\enquote {\bibinfo {title} {{Branching of quasinormal modes for nearly
  extremal Kerr black holes}},}\ }\href {\doibase 10.1103/PhysRevD.87.041502}
  {\bibfield  {journal} {\bibinfo  {journal} {Phys. Rev. D}\ }\textbf {\bibinfo
  {volume} {87}},\ \bibinfo {pages} {041502} (\bibinfo {year}
  {2013}{\natexlab{a}})},\ \Eprint {http://arxiv.org/abs/1212.3271}
  {arXiv:1212.3271 [gr-qc]} \BibitemShut {NoStop}%
\bibitem [{\citenamefont {Yang}\ \emph
  {et~al.}(2013{\natexlab{b}})\citenamefont {Yang}, \citenamefont {Zimmerman},
  \citenamefont {Zengino\u{g}lu}, \citenamefont {Zhang}, \citenamefont
  {Berti},\ and\ \citenamefont {Chen}}]{Yang:2013uba}%
  \BibitemOpen
  \bibfield  {author} {\bibinfo {author} {\bibfnamefont {Huan}\ \bibnamefont
  {Yang}}, \bibinfo {author} {\bibfnamefont {Aaron}\ \bibnamefont {Zimmerman}},
  \bibinfo {author} {\bibfnamefont {An\i{}l}\ \bibnamefont {Zengino\u{g}lu}},
  \bibinfo {author} {\bibfnamefont {Fan}\ \bibnamefont {Zhang}}, \bibinfo
  {author} {\bibfnamefont {Emanuele}\ \bibnamefont {Berti}}, \ and\ \bibinfo
  {author} {\bibfnamefont {Yanbei}\ \bibnamefont {Chen}},\ }\bibfield  {title}
  {\enquote {\bibinfo {title} {{Quasinormal modes of nearly extremal Kerr
  spacetimes: spectrum bifurcation and power-law ringdown}},}\ }\href {\doibase
  10.1103/PhysRevD.88.044047} {\bibfield  {journal} {\bibinfo  {journal} {Phys.
  Rev. D}\ }\textbf {\bibinfo {volume} {88}},\ \bibinfo {pages} {044047}
  (\bibinfo {year} {2013}{\natexlab{b}})},\ \Eprint
  {http://arxiv.org/abs/1307.8086} {arXiv:1307.8086 [gr-qc]} \BibitemShut
  {NoStop}%
\bibitem [{\citenamefont {Astefanesei}\ \emph {et~al.}(2006)\citenamefont
  {Astefanesei}, \citenamefont {Goldstein}, \citenamefont {Jena}, \citenamefont
  {Sen},\ and\ \citenamefont {Trivedi}}]{Astefanesei:2006dd}%
  \BibitemOpen
  \bibfield  {author} {\bibinfo {author} {\bibfnamefont {Dumitru}\ \bibnamefont
  {Astefanesei}}, \bibinfo {author} {\bibfnamefont {Kevin}\ \bibnamefont
  {Goldstein}}, \bibinfo {author} {\bibfnamefont {Rudra~P.}\ \bibnamefont
  {Jena}}, \bibinfo {author} {\bibfnamefont {Ashoke}\ \bibnamefont {Sen}}, \
  and\ \bibinfo {author} {\bibfnamefont {Sandip~P.}\ \bibnamefont {Trivedi}},\
  }\bibfield  {title} {\enquote {\bibinfo {title} {{Rotating attractors}},}\
  }\href {\doibase 10.1088/1126-6708/2006/10/058} {\bibfield  {journal}
  {\bibinfo  {journal} {JHEP}\ }\textbf {\bibinfo {volume} {10}},\ \bibinfo
  {pages} {058} (\bibinfo {year} {2006})},\ \Eprint
  {http://arxiv.org/abs/hep-th/0606244} {arXiv:hep-th/0606244} \BibitemShut
  {NoStop}%
\bibitem [{\citenamefont {Gralla}\ and\ \citenamefont
  {Zimmerman}(2018)}]{Gralla:2018xzo}%
  \BibitemOpen
  \bibfield  {author} {\bibinfo {author} {\bibfnamefont {Samuel~E.}\
  \bibnamefont {Gralla}}\ and\ \bibinfo {author} {\bibfnamefont {Peter}\
  \bibnamefont {Zimmerman}},\ }\bibfield  {title} {\enquote {\bibinfo {title}
  {{Scaling and Universality in Extremal Black Hole Perturbations}},}\ }\href
  {\doibase 10.1007/JHEP06(2018)061} {\bibfield  {journal} {\bibinfo  {journal}
  {JHEP}\ }\textbf {\bibinfo {volume} {06}},\ \bibinfo {pages} {061} (\bibinfo
  {year} {2018})},\ \Eprint {http://arxiv.org/abs/1804.04753} {arXiv:1804.04753
  [gr-qc]} \BibitemShut {NoStop}%
\bibitem [{\citenamefont {Olver}\ \emph {et~al.}(2016)\citenamefont {Olver},
  \citenamefont {Daalhuis}, \citenamefont {Lozier}, \citenamefont {Schneider},
  \citenamefont {Boisvert}, \citenamefont {Clark}, \citenamefont {Miller},\
  and\ \citenamefont {Saunders}}]{olver2016nist}%
  \BibitemOpen
  \bibfield  {author} {\bibinfo {author} {\bibfnamefont {FWJ}\ \bibnamefont
  {Olver}}, \bibinfo {author} {\bibfnamefont {AB~Olde}\ \bibnamefont
  {Daalhuis}}, \bibinfo {author} {\bibfnamefont {DW}~\bibnamefont {Lozier}},
  \bibinfo {author} {\bibfnamefont {BI}~\bibnamefont {Schneider}}, \bibinfo
  {author} {\bibfnamefont {RF}~\bibnamefont {Boisvert}}, \bibinfo {author}
  {\bibfnamefont {CW}~\bibnamefont {Clark}}, \bibinfo {author} {\bibfnamefont
  {BR}~\bibnamefont {Miller}}, \ and\ \bibinfo {author} {\bibfnamefont
  {BV}~\bibnamefont {Saunders}},\ }\bibfield  {title} {\enquote {\bibinfo
  {title} {Nist digital library of mathematical functions http://dlmf. nist.
  gov},}\ }\href@noop {} {\bibfield  {journal} {\bibinfo  {journal} {Release}\
  }\textbf {\bibinfo {volume} {1}},\ \bibinfo {pages} {22} (\bibinfo {year}
  {2016})}\BibitemShut {NoStop}%
\bibitem [{\citenamefont {Lechner}\ \emph {et~al.}(2002)\citenamefont
  {Lechner}, \citenamefont {Thornburg}, \citenamefont {Husa},\ and\
  \citenamefont {Aichelburg}}]{Lechner:2001ng}%
  \BibitemOpen
  \bibfield  {author} {\bibinfo {author} {\bibfnamefont {Christiane}\
  \bibnamefont {Lechner}}, \bibinfo {author} {\bibfnamefont {Jonathan}\
  \bibnamefont {Thornburg}}, \bibinfo {author} {\bibfnamefont {Sascha}\
  \bibnamefont {Husa}}, \ and\ \bibinfo {author} {\bibfnamefont {Peter~C.}\
  \bibnamefont {Aichelburg}},\ }\bibfield  {title} {\enquote {\bibinfo {title}
  {{A New transition between discrete and continuous selfsimilarity in critical
  gravitational collapse}},}\ }\href {\doibase 10.1103/PhysRevD.65.081501}
  {\bibfield  {journal} {\bibinfo  {journal} {Phys. Rev. D}\ }\textbf {\bibinfo
  {volume} {65}},\ \bibinfo {pages} {081501} (\bibinfo {year} {2002})},\
  \Eprint {http://arxiv.org/abs/gr-qc/0112008} {arXiv:gr-qc/0112008}
  \BibitemShut {NoStop}%
\bibitem [{\citenamefont {Cardy}(1996)}]{Cardy1996}%
  \BibitemOpen
  \bibfield  {author} {\bibinfo {author} {\bibfnamefont {John}\ \bibnamefont
  {Cardy}},\ }\href {\doibase 10.1017/CBO9781316036440} {\emph {\bibinfo
  {title} {Scaling and Renormalization in Statistical Physics}}},\ Cambridge
  Lecture Notes in Physics\ (\bibinfo  {publisher} {Cambridge University
  Press},\ \bibinfo {year} {1996})\BibitemShut {NoStop}%
\bibitem [{\citenamefont {Sachdev}(2011)}]{Sachdev2011}%
  \BibitemOpen
  \bibfield  {author} {\bibinfo {author} {\bibfnamefont {Subir}\ \bibnamefont
  {Sachdev}},\ }\href {\doibase 10.1017/CBO9780511973765} {\emph {\bibinfo
  {title} {Quantum Phase Transitions}}},\ \bibinfo {edition} {2nd}\ ed.\
  (\bibinfo  {publisher} {Cambridge University Press},\ \bibinfo {year}
  {2011})\BibitemShut {NoStop}%
\bibitem [{\citenamefont {{Kaplan}}\ \emph {et~al.}(2009)\citenamefont
  {{Kaplan}}, \citenamefont {{Lee}}, \citenamefont {{Son}},\ and\ \citenamefont
  {{Stephanov}}}]{Kaplan2009}%
  \BibitemOpen
  \bibfield  {author} {\bibinfo {author} {\bibfnamefont {David~B.}\
  \bibnamefont {{Kaplan}}}, \bibinfo {author} {\bibfnamefont {Jong-Wan}\
  \bibnamefont {{Lee}}}, \bibinfo {author} {\bibfnamefont {Dam~T.}\
  \bibnamefont {{Son}}}, \ and\ \bibinfo {author} {\bibfnamefont {Mikhail~A.}\
  \bibnamefont {{Stephanov}}},\ }\bibfield  {title} {\enquote {\bibinfo {title}
  {{Conformality lost}},}\ }\href {\doibase 10.1103/PhysRevD.80.125005}
  {\bibfield  {journal} {\bibinfo  {journal} {\prd}\ }\textbf {\bibinfo
  {volume} {80}},\ \bibinfo {eid} {125005} (\bibinfo {year} {2009})},\ \Eprint
  {http://arxiv.org/abs/0905.4752} {arXiv:0905.4752 [hep-th]} \BibitemShut
  {NoStop}%
\bibitem [{\citenamefont {Peskin}\ and\ \citenamefont
  {Schroeder}(1995)}]{Peskin1995}%
  \BibitemOpen
  \bibfield  {author} {\bibinfo {author} {\bibfnamefont {Michael~E.}\
  \bibnamefont {Peskin}}\ and\ \bibinfo {author} {\bibfnamefont {Daniel~V.}\
  \bibnamefont {Schroeder}},\ }\href {\doibase 10.1201/9780429503559} {\emph
  {\bibinfo {title} {{An Introduction to quantum field theory}}}}\ (\bibinfo
  {publisher} {Addison-Wesley},\ \bibinfo {address} {Reading, USA},\ \bibinfo
  {year} {1995})\BibitemShut {NoStop}%
\bibitem [{\citenamefont {Gang}(2007)}]{Wen2004}%
  \BibitemOpen
  \bibfield  {author} {\bibinfo {author} {\bibfnamefont {Wen~Xiao}\
  \bibnamefont {Gang}},\ }\href {\doibase
  10.1093/acprof:oso/9780199227259.001.0001} {\emph {\bibinfo {title} {{Quantum
  field theory of many-body systems: from the origin of sound to an origin of
  light and electrons}}}}\ (\bibinfo  {publisher} {Oxford University Press},\
  \bibinfo {address} {Oxford},\ \bibinfo {year} {2007})\BibitemShut {NoStop}%
\bibitem [{\citenamefont {Wilson}(1971)}]{Wilson1971}%
  \BibitemOpen
  \bibfield  {author} {\bibinfo {author} {\bibfnamefont {Kenneth~G.}\
  \bibnamefont {Wilson}},\ }\bibfield  {title} {\enquote {\bibinfo {title}
  {Renormalization group and strong interactions},}\ }\href {\doibase
  10.1103/PhysRevD.3.1818} {\bibfield  {journal} {\bibinfo  {journal} {Phys.
  Rev. D}\ }\textbf {\bibinfo {volume} {3}},\ \bibinfo {pages} {1818--1846}
  (\bibinfo {year} {1971})}\BibitemShut {NoStop}%
\bibitem [{\citenamefont {{Bernard}}\ and\ \citenamefont
  {{LeClair}}(2001)}]{Bernard2001}%
  \BibitemOpen
  \bibfield  {author} {\bibinfo {author} {\bibfnamefont {D.}~\bibnamefont
  {{Bernard}}}\ and\ \bibinfo {author} {\bibfnamefont {A.}~\bibnamefont
  {{LeClair}}},\ }\bibfield  {title} {\enquote {\bibinfo {title} {{Strong-weak
  coupling duality in anisotropic current interactions}},}\ }\href {\doibase
  10.1016/S0370-2693(01)00695-5} {\bibfield  {journal} {\bibinfo  {journal}
  {Physics Letters B}\ }\textbf {\bibinfo {volume} {512}},\ \bibinfo {pages}
  {78--84} (\bibinfo {year} {2001})},\ \Eprint
  {http://arxiv.org/abs/hep-th/0103096} {arXiv:hep-th/0103096 [hep-th]}
  \BibitemShut {NoStop}%
\bibitem [{\citenamefont {{LeClair}}\ \emph {et~al.}(2003)\citenamefont
  {{LeClair}}, \citenamefont {{Rom{\'a}n}},\ and\ \citenamefont
  {{Sierra}}}]{LeClair2003a}%
  \BibitemOpen
  \bibfield  {author} {\bibinfo {author} {\bibfnamefont {Andr{\'e}}\
  \bibnamefont {{LeClair}}}, \bibinfo {author} {\bibfnamefont
  {Jos{\'e}~Mar{\'\i}a.}\ \bibnamefont {{Rom{\'a}n}}}, \ and\ \bibinfo {author}
  {\bibfnamefont {Germ{\'a}n}\ \bibnamefont {{Sierra}}},\ }\bibfield  {title}
  {\enquote {\bibinfo {title} {{Russian doll renormalization group and
  Kosterlitz-Thouless flows}},}\ }\href {\doibase
  10.1016/j.nuclphysb.2003.09.032} {\bibfield  {journal} {\bibinfo  {journal}
  {Nuclear Physics B}\ }\textbf {\bibinfo {volume} {675}},\ \bibinfo {pages}
  {584--606} (\bibinfo {year} {2003})},\ \Eprint
  {http://arxiv.org/abs/hep-th/0301042} {arXiv:hep-th/0301042 [hep-th]}
  \BibitemShut {NoStop}%
\bibitem [{\citenamefont {{LeClair}}\ \emph {et~al.}(2004)\citenamefont
  {{LeClair}}, \citenamefont {{Rom{\'a}n}},\ and\ \citenamefont
  {{Sierra}}}]{LeClair2003}%
  \BibitemOpen
  \bibfield  {author} {\bibinfo {author} {\bibfnamefont {Andr{\'e}}\
  \bibnamefont {{LeClair}}}, \bibinfo {author} {\bibfnamefont
  {Jos{\'e}~Mar{\'\i}a}\ \bibnamefont {{Rom{\'a}n}}}, \ and\ \bibinfo {author}
  {\bibfnamefont {Germ{\'a}n}\ \bibnamefont {{Sierra}}},\ }\bibfield  {title}
  {\enquote {\bibinfo {title} {{Log-periodic behavior of finite size effects in
  field theories with RG limit cycles}},}\ }\href {\doibase
  10.1016/j.nuclphysb.2004.08.033} {\bibfield  {journal} {\bibinfo  {journal}
  {Nuclear Physics B}\ }\textbf {\bibinfo {volume} {700}},\ \bibinfo {pages}
  {407--435} (\bibinfo {year} {2004})},\ \Eprint
  {http://arxiv.org/abs/hep-th/0312141} {arXiv:hep-th/0312141 [hep-th]}
  \BibitemShut {NoStop}%
\bibitem [{\citenamefont {{Jepsen}}\ \emph {et~al.}(2021)\citenamefont
  {{Jepsen}}, \citenamefont {{Klebanov}},\ and\ \citenamefont
  {{Popov}}}]{Jepsen2020}%
  \BibitemOpen
  \bibfield  {author} {\bibinfo {author} {\bibfnamefont {Christian~B.}\
  \bibnamefont {{Jepsen}}}, \bibinfo {author} {\bibfnamefont {Igor~R.}\
  \bibnamefont {{Klebanov}}}, \ and\ \bibinfo {author} {\bibfnamefont
  {Fedor~K.}\ \bibnamefont {{Popov}}},\ }\bibfield  {title} {\enquote {\bibinfo
  {title} {{RG limit cycles and unconventional fixed points in perturbative
  QFT}},}\ }\href {\doibase 10.1103/PhysRevD.103.046015} {\bibfield  {journal}
  {\bibinfo  {journal} {\prd}\ }\textbf {\bibinfo {volume} {103}},\ \bibinfo
  {eid} {046015} (\bibinfo {year} {2021})},\ \Eprint
  {http://arxiv.org/abs/2010.15133} {arXiv:2010.15133 [hep-th]} \BibitemShut
  {NoStop}%
\bibitem [{\citenamefont {Huse}(1981)}]{Huse1991}%
  \BibitemOpen
  \bibfield  {author} {\bibinfo {author} {\bibfnamefont {David~A.}\
  \bibnamefont {Huse}},\ }\bibfield  {title} {\enquote {\bibinfo {title}
  {Simple three-state model with infinitely many phases},}\ }\href {\doibase
  10.1103/PhysRevB.24.5180} {\bibfield  {journal} {\bibinfo  {journal} {Phys.
  Rev. B}\ }\textbf {\bibinfo {volume} {24}},\ \bibinfo {pages} {5180--5194}
  (\bibinfo {year} {1981})}\BibitemShut {NoStop}%
\bibitem [{\citenamefont {Veytsman}(1993)}]{Veytsman1993}%
  \BibitemOpen
  \bibfield  {author} {\bibinfo {author} {\bibfnamefont {Boris~A.}\
  \bibnamefont {Veytsman}},\ }\bibfield  {title} {\enquote {\bibinfo {title}
  {Limit cycles in renormalization group flows: thermodynamics controls dances
  of space patterns},}\ }\href {\doibase
  https://doi.org/10.1016/0375-9601(93)90463-A} {\bibfield  {journal} {\bibinfo
   {journal} {Physics Letters A}\ }\textbf {\bibinfo {volume} {183}},\ \bibinfo
  {pages} {315--318} (\bibinfo {year} {1993})}\BibitemShut {NoStop}%
\bibitem [{\citenamefont {{Bedaque}}\ \emph {et~al.}(1999)\citenamefont
  {{Bedaque}}, \citenamefont {{Hammer}},\ and\ \citenamefont {{van
  Kolck}}}]{Bedaque1998}%
  \BibitemOpen
  \bibfield  {author} {\bibinfo {author} {\bibfnamefont {P.~F.}\ \bibnamefont
  {{Bedaque}}}, \bibinfo {author} {\bibfnamefont {H.~W.}\ \bibnamefont
  {{Hammer}}}, \ and\ \bibinfo {author} {\bibfnamefont {U.}~\bibnamefont {{van
  Kolck}}},\ }\bibfield  {title} {\enquote {\bibinfo {title} {{Renormalization
  of the Three-Body System with Short-Range Interactions}},}\ }\href {\doibase
  10.1103/PhysRevLett.82.463} {\bibfield  {journal} {\bibinfo  {journal}
  {\prl}\ }\textbf {\bibinfo {volume} {82}},\ \bibinfo {pages} {463--467}
  (\bibinfo {year} {1999})},\ \Eprint {http://arxiv.org/abs/nucl-th/9809025}
  {arXiv:nucl-th/9809025 [nucl-th]} \BibitemShut {NoStop}%
\bibitem [{\citenamefont {{G{\l}azek}}\ and\ \citenamefont
  {{Wilson}}(2002)}]{Glazek2002}%
  \BibitemOpen
  \bibfield  {author} {\bibinfo {author} {\bibfnamefont {Stanis{\l}aw~D.}\
  \bibnamefont {{G{\l}azek}}}\ and\ \bibinfo {author} {\bibfnamefont
  {Kenneth~G.}\ \bibnamefont {{Wilson}}},\ }\bibfield  {title} {\enquote
  {\bibinfo {title} {{Limit Cycles in Quantum Theories}},}\ }\href {\doibase
  10.1103/PhysRevLett.89.230401} {\bibfield  {journal} {\bibinfo  {journal}
  {\prl}\ }\textbf {\bibinfo {volume} {89}},\ \bibinfo {eid} {230401} (\bibinfo
  {year} {2002})},\ \Eprint {http://arxiv.org/abs/hep-th/0203088}
  {arXiv:hep-th/0203088 [hep-th]} \BibitemShut {NoStop}%
\bibitem [{\citenamefont {{Hartnoll}}\ \emph {et~al.}(2016)\citenamefont
  {{Hartnoll}}, \citenamefont {{Ramirez}},\ and\ \citenamefont
  {{Santos}}}]{Hartnoll2015}%
  \BibitemOpen
  \bibfield  {author} {\bibinfo {author} {\bibfnamefont {Sean~A.}\ \bibnamefont
  {{Hartnoll}}}, \bibinfo {author} {\bibfnamefont {David~M.}\ \bibnamefont
  {{Ramirez}}}, \ and\ \bibinfo {author} {\bibfnamefont {Jorge~E.}\
  \bibnamefont {{Santos}}},\ }\bibfield  {title} {\enquote {\bibinfo {title}
  {{Thermal conductivity at a disordered quantum critical point}},}\ }\href
  {\doibase 10.1007/JHEP04(2016)022} {\bibfield  {journal} {\bibinfo  {journal}
  {Journal of High Energy Physics}\ }\textbf {\bibinfo {volume} {2016}},\
  \bibinfo {eid} {22} (\bibinfo {year} {2016})},\ \Eprint
  {http://arxiv.org/abs/1508.04435} {arXiv:1508.04435 [hep-th]} \BibitemShut
  {NoStop}%
\bibitem [{\citenamefont {{Yerzhakov}}\ and\ \citenamefont
  {{Maciejko}}(2021)}]{Yerzhakov2020}%
  \BibitemOpen
  \bibfield  {author} {\bibinfo {author} {\bibfnamefont {Hennadii}\
  \bibnamefont {{Yerzhakov}}}\ and\ \bibinfo {author} {\bibfnamefont {Joseph}\
  \bibnamefont {{Maciejko}}},\ }\bibfield  {title} {\enquote {\bibinfo {title}
  {{Random-mass disorder in the critical Gross-Neveu-Yukawa models}},}\ }\href
  {\doibase 10.1016/j.nuclphysb.2020.115241} {\bibfield  {journal} {\bibinfo
  {journal} {Nuclear Physics B}\ }\textbf {\bibinfo {volume} {962}},\ \bibinfo
  {eid} {115241} (\bibinfo {year} {2021})},\ \Eprint
  {http://arxiv.org/abs/2008.13663} {arXiv:2008.13663 [cond-mat.str-el]}
  \BibitemShut {NoStop}%
\bibitem [{\citenamefont {{Bulycheva}}\ and\ \citenamefont
  {{Gorsky}}(2014)}]{Bulycheva2014}%
  \BibitemOpen
  \bibfield  {author} {\bibinfo {author} {\bibfnamefont {K.}~\bibnamefont
  {{Bulycheva}}}\ and\ \bibinfo {author} {\bibfnamefont {A.}~\bibnamefont
  {{Gorsky}}},\ }\bibfield  {title} {\enquote {\bibinfo {title} {{Rg Limit
  Cycles}},}\ }in\ \href {\doibase 10.1142/9789814616850\_0005} {\emph
  {\bibinfo {booktitle} {Pomeranchuk 100}}},\ \bibinfo {editor} {edited by\
  \bibinfo {editor} {\bibfnamefont {Alexander~S.}\ \bibnamefont {{Gorsky}}}\
  and\ \bibinfo {editor} {\bibfnamefont {I.}~\bibnamefont {{Vysotsky
  Mikhail}}}}\ (\bibinfo {year} {2014})\ pp.\ \bibinfo {pages} {82--112},\
  \Eprint {http://arxiv.org/abs/1402.2431} {arXiv:1402.2431 [hep-th]}
  \BibitemShut {NoStop}%
\end{thebibliography}%

\end{document}